\documentclass[%
 reprint,
superscriptaddress,
 amsmath,amssymb,
 aps,
]{revtex4-2}

\usepackage{graphicx}
\usepackage{dcolumn}
\usepackage{bm}
\usepackage{xcolor}

\begin{document}
\title{Robust Identification of Topological Phase Transition by Self-Supervised Machine Learning Approach}

\author{Chi-Ting Ho}
\affiliation{Physics Division, National Center for Theoretical Sciences, Hsinchu 30013, Taiwan}
\affiliation{Physics Department, National Tsing Hua University, Hsinchu 30013, Taiwan}

\author{Daw-Wei Wang}
\affiliation{Physics Division, National Center for Theoretical Sciences, Hsinchu 30013, Taiwan}
\affiliation{Physics Department, National Tsing Hua University, Hsinchu 30013, Taiwan}
\affiliation{Center for Theory and Computation, National Tsing Hua University, Hsinchu 30013, Taiwan}
\affiliation{Center for Quantum Technology, National Tsing Hua University, Hsinchu 30013, Taiwan}


\begin{abstract}
We propose a systematic methodology to identify the topological phase transition through a self-supervised machine learning model, which is trained to correlate system parameters to the non-local observables in time-of-flight experiments of ultracold atoms. Different from the conventional supervised learning approach, where the predicted phase transition point is very sensitive to the training region and data labelling, our self-supervised learning approach identifies the phase transition point by the largest deviation of the predicted results from the known system parameters and by the highest confidence through a systematic shift of the training regions. We demonstrate the robust application of this approach results in various 1D and 2D exactly solvable models, using different input features (time-of-flight images, spatial correlation function or density-density correlation function). As a result, our self-supervised approach should be a very general and reliable method for many condensed matter or solid state systems to observe new states of matters solely based on experimental measurements, even without a priori knowledge of the phase transition models.
\end{abstract}

\maketitle

\section{Introduction}
\label{sec:introduction}

Identifying many-body phase boundaries in the phase diagram is one of the most important subjects in condensed matter physics, no matter for classical phase transitions, quantum phase transitions, or topological phase transitions. From theoretical perspectives, these tasks are usually done by analytically solving simpler models in thermodynamic limit or by numerically calculating physical quantities of a finite system. It is expected that the obtained physical quantities, say ground state energy, order parameters, or entanglement entropy, show some singular behaviors near the phase transition boundary if the system size is large enough \cite{Andersonbook}. 

From experimental perspectives, however, it is much more challenging to identify a topological phase transition than to identify classical or quantum phase transitions, because, the former has no order parameter or divergent thermodynamic quantities at the phase boundary. The identifiers of topological properties, say Chern number, winding number, entanglement entropy, or entanglement spectrum \cite{Intro_Chern_1, Intro_Chern_2, Topo_book, calu_Entangle, Intro_Entangle} etc., are all non-local quantities and hence not easily observed in ordinary experimental measurements. Most experimentally observed topological phases are based on the transport properties of the edge states \cite{TI_transport_1, TI_transport_2}, which originate from the topological band structure in the bulk. However, such single particle measurement could be very challenging in the systems of ultracold atoms due to the large density fluctuations and the harmonic trapping potential. Most proposals for measuring the topological states in ultracold atoms depend on the specific dynamical measurement and hence cannot be unambiguously applied in other situations \cite{TI_ultracold_1, TI_ultracold_2, TI_ultracold_3, TI_ultracold_4, TI_ultracold_5}.

Recent rapid development of Artificial Intelligence (AI) and Machine Learning (ML) methods open new perspectives on these issues. One of the most popular application of ML is to utilize Supervised Learning (SL) approach for the identification of classical \cite{SL_ClassicalIsing, SL_ClassicalXY} or quantum phase transitions \cite{SL_FermiHubbard, SL_Kitaev, SL_Haldane, SL_BoseHubbard}. The basic scenario is to treat many-body phases as the objects to be identified in Computer Vision (CV) problems. However, the validity of such approach strongly depends on known results in the training data, while this information is usually not available in a new system or an unsolved model. On the other hand, the applications of Unsupervised Learning (USL) approach, such as Principal Component Analysis (PCA) \cite{PCA}, K-means clustering \cite{k_means} and Variational Autoencoder (VAE) \cite{VAE} etc. are also developed for the identification of phase transitions \cite{PCA_Ising,USL_ClassicalXY,USL_Hubbard,USL_Ising, USL_ANNNI}. 
However, these unsupervised learning approaches could not provide a precise determination of the phase boundary of topological matters, because the topological edge states could be easily overwhelmed by the bulk state properties.

In this paper, we propose a self-supervised learning (SSL) approach to provide a much more robust and general methodology for the identification of topological phase transition. We first train a model to simulate the internal relationship between non-local experimental observables and system parameters, where the latter are controlled independently by the experimental setup. The candidates of a phase transition point (PTP) is then defined by measuring the largest deviation between the predicted results and the actual system parameters in the test region. Finally, we shift the training/test regions systematically and the true PTP is then identified by the one with the highest confidence (defined below), indicating how much one could trust this predicted results. 

To demonstrate the effectiveness and generality of our SSL approach, we calculate four different types of topological models, including 1D Su-Schrieffer–Heeger model, 2D Haldane model, 1D Kitaev chain model, and 2D chiral superfluid model, and show how a precise PTP could be obtained with very high confidence level. This is because a topological phase transition could be understood a singularity of a "microscopic state function", which contains the information of system entanglement. Similar conclusion can be also obtained by using spatial correlation function and density-density correlation function as the input features, showing that our approach could be easily generalized to other condensed matter systems with different experimental measurements. We emphasize that these results are solely based on the experimental data even without assuming the existence of these phase transitions and/or any theoretical understanding of the phases, and therefore could be applied to find other new phases of matter in the future experiments. 

The structure of this paper is following: In Sec. \ref{sec:supervised approach}, we discuss the problems of the conventional supervised learning approach when applied to the phase transition identification in condensed matter physics, and show why most results obtained so far may not be reliable after more careful examination. In Sec. \ref{sec:self-supervised learning}, we introduce a self-supervised learning approach to resolve the present problems and show how a new methodology could be operated. We then apply this approach to 1D SSH model and 2D Haldane model as examples of intrinsic topological phases in Sec. \ref{sec:Intrisic Topological}, and then to 1D Kitaev model and 2D $p$-wave model as examples of symmetry protected topological phase in Sec. \ref{sec:Symmetry Topological}. In Sec. \ref{sec:discussion}, we then discuss the results using spatial correlation function and density-density correlation function as input features, together with the finite trapping potential effects in the ToF images, demonstrating the generality and robustness of our SSL method. Finally, we conclude our work in Sec. \ref{sec:conclusion}. More details of ToF image calculation and machine learning parameters are provided in Appendix \ref{sec:input features} and \ref{sec:ML parameters} respectively.

\section{Conventional Supervised Learning Approaches and Related Problems}
\label{sec:supervised approach}

Supervised Learning (SL) is the most popular machine learning method in different applications, because the model is trained by numerous labeled data and hence can provide accurate prediction for the test data \cite{ML_book}. Take computer vision (CV) as an example, thousands of objects could be accurately identified by using a Convolutional Neural Network (CNN) after trained by millions of images \cite{AlexNet, ResNet}. Since different many-body phases could be understood as different categories of objects, it is reasonable to expect that these phases could be also classified via a SL method, where the input features are experimentally/numerically obtained quantities for a given system parameter. However, there are two fundamental problems for such a naive extension in condensed matter physics, as described in details below. 

\begin{figure}[h]
    \centering
    \includegraphics[width=0.48\textwidth]{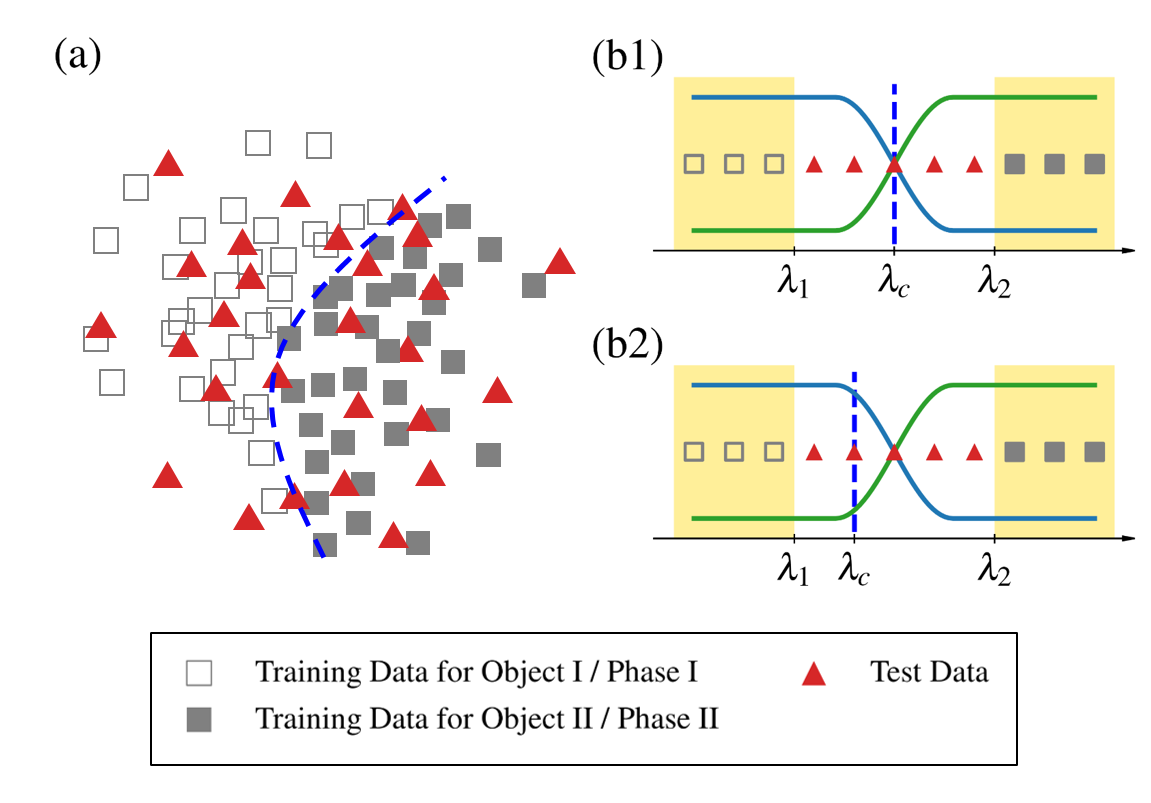}
    \caption{
    (a) shows a schematic data distribution of a typical CV problem, obtained by projecting the original high-dimensional feature space onto a two-dimensional effective space through tSNE method (see Ref. \cite{tSNE}). 
    The training data of these two objects have to distribute closely in space so that the hidden boundary (blue dashed line) could be identified by the predicted results of test data.
    (b1) and (b2) show schematic data distribution of a conventional SL approach for the identification of the PTP with two different training regions (defined by $\lambda_1$ and $\lambda_2$ here with yellow background for simplicity). The test data are in the middle region ($\lambda\in (\lambda_1,\lambda_2)$) with the blue and green curves standing for the probabilities predicted to be Phase I and Phase II respectively. The crossing point with probability 0.5 is interpreted as the predicted PTP (see the text). However, the crossing point can be strongly influenced by the training region so that the predicted PTP may be deviated from its true value ($\lambda_c$, vertical dashed lines), as shown in (b2).
    }
    \label{fig:scheme_distribution.png}
\end{figure}

\subsection{Problem I: Data Distribution}

In order to clarify these problems, we first show a schematic distribution of the training data and the test data of a typical CV task in Fig. \ref{fig:scheme_distribution.png}(a). These data points are originally distributed in a much higher dimensional feature space, but can be effectively expressed in a two-dimensional space through t-distributed stochastic neighbor embedding (tSNE) method \cite{tSNE}. One could see that although the training data of the two objects (Object A and Object B) are distributed in different regions, the test data are also in similar region, so that the "boundary" to classify these two objects could then be obtained as the dashed line by equating the probabilities to be Object A and Object B. However, we have to note that such a "boundary" should exist only in the original higher dimensional feature space, and here we just show an effective representation for a simpler comparison to phase identification problems below. 

When applying similar SL approach to identify the PTP in condensed matter systems \cite{SL_ClassicalIsing, SL_ClassicalXY,SL_FermiHubbard, SL_Kitaev, SL_Haldane, SL_BoseHubbard}, however, the data distribution has a very different pattern from CV task, as shown in Figs. \ref{fig:scheme_distribution.png}(b1) and (b2). The training data of these two phases are usually separated away from each other in order to unambiguously define their phases: the training data with $\lambda<\lambda_1$ are labeled as Phase A, the data training with $\lambda>\lambda_2$ are labeled as Phase B with $\lambda_1<\lambda_2$. As a result, although one may still define the PTP by finding the position of equal probabilities to be these two phases ($P_A=P_B=0.5$), the associate confidence should be very weak, because, different from the CV task shown in (a), there is NO training data with known labels distributed near the PTP. In other words, it is not convincing why the obtained PTP captures any realistic and reliable physical properties for the determination of the phase transition.

\subsection{Problem II: Training Region Dependence}

Besides the data distribution issues, the second problem of conventional SL approach for the identification of phase transitions is that one has to assume the existence of these two phases in the test region. In other words, it could still provide similar results even there is no such a phase transition in the parameter region. This issue could be understood by using different training/test regions, as shown in Figs. \ref{fig:scheme_distribution.png}(b1) and (b2). One could see that even the two phases have been correctly labeled in the training data (Phase A for $\lambda<\lambda_1$ and Phase B for $\lambda>\lambda_B$), the predicted position of the PTP can be strongly depend on the choices of training region, i.e. $\lambda_1$ and $\lambda_2$. Furthermore, as we will see below, similar situation occurs even there is \textit{no} phase transition inside the test region. Therefore, the conventional SL approach could not provide a reliable prediction of the PTP without \textit{a priori} knowledge about its position. One could not use this approach to find any new phase transition.

\section{Self-Supervised Learning Approach} 
\label{sec:self-supervised learning}

\subsection{Conjecture: Simulating the Microscopic State Function of a Many-Body System}

In order to resolve the two problems mentioned above for the conventional SL approach in the identification of the PTP, here we propose a self-supervised learning (SSL) approach by using the same data distribution, but trained and analyzed totally differently: When the experimentally observed features (say ToF images, see Appendix \ref{sec:input features} for details) of each data are input, the SSL model is trained not by artificial labels (say Phase A/B), but by the associated system parameter, $\lambda$ (see Fig. \ref{fig:scheme_SSL.png}(a)). It is our theoretical conjecture that such a self-training scheme could simulate the internal connection between the input features (here denoted by $\{x\}$) and the output value, $\lambda$.

More precisely, we define such a theoretical functional relationship, $F_{MSF}(\{x\},\lambda)=0$, to be a Microscopic State Function (MSF) of a many-body system, which is similar to the State Function known in Thermodynamics or Statistical Mechanics \cite{Pathriabook}, a theoretical functional relationship between macroscopic quantities (say total energy, total entropy, total particle number etc.) only. The MSF we defined above contains correlation functions between various \textit{microscopic} physical quantities, say the momentum distribution, the spatial two-point correlation function in the ToF images, and/or the density-density correlation function, depending on the input features from the experiments. As a result, the information of many-body entanglement should be also embedded inside such a MSF through an implicit functional form. Although such a complicated functional relationship may not be derivable theoretically, it could be still simulated by machine learning method through our SSL approach. For example, when using ToF images as the input features for our SSL approach (see Appendix \ref{sec:input features}), we have
\begin{eqnarray}
F_{SSL}(n_{ToF}(x,y),\lambda)\to F_{MSF}(\{\langle \hat{c}^\dagger_i \hat{c}^{}_j\rangle\},\lambda),
\end{eqnarray}
for a successful simulation in the \textit{training region}, i.e. $\lambda<\lambda_1$ or $\lambda>\lambda_2$ in Fig. \ref{fig:scheme_SSL.png}(b) and (c). Here $\hat{c}_i$ is the field operator of the associated particles.

Above approximation provided by the SSL approach makes it possible to identify topological phase transitions by measuring the singular point of the simulated MSF, just as measuring the singularity of entanglement entropy for a topological phase transition. However, since the explicit form of $F_{MSF}(\{\langle \hat{c}^\dagger_i \hat{c}^{}_j\rangle \},\lambda)$ is unknown, such singularity has to be obtained by measuring the position, $\lambda_t$, where the simulation fails most in the \textit{test region}, i.e. $\lambda\in(\lambda_1,\lambda_2)$ in Fig. \ref{fig:scheme_SSL.png}(b) and (c).
Since one may always find a $\lambda_t$ no matter if a true PTP is involved inside the test region, it is therefore reasonable to expect that the obtained $\lambda_t$ should be pinned at $\lambda_c$, if the latter is actually included in the test region, no matter how the training region changes. Different from the calculation of entanglement entropy, the MSF simulated by our SSL approach can be obtained via experimentally measurable quantities (say the ToF images), while the entanglement entropy could not be directly measured in most experiments. This is the theoretical conjecture behind our SSL approach for the identification of topological phase transition, but we admit that a more concrete theory will be needed for more quantitative studies in the future. Below we will just describe how to extract the signatures of topological phase transitions from data analysis point of view, which could be applied even without assuming their presence or any theoretical understanding beforehand.

\subsection{Signature of Phase Transitions (I): The Largest Deviation of the Prediction from the Actual System Parameters}

\begin{figure}[h]
    \centering
    \includegraphics[width=0.48\textwidth]{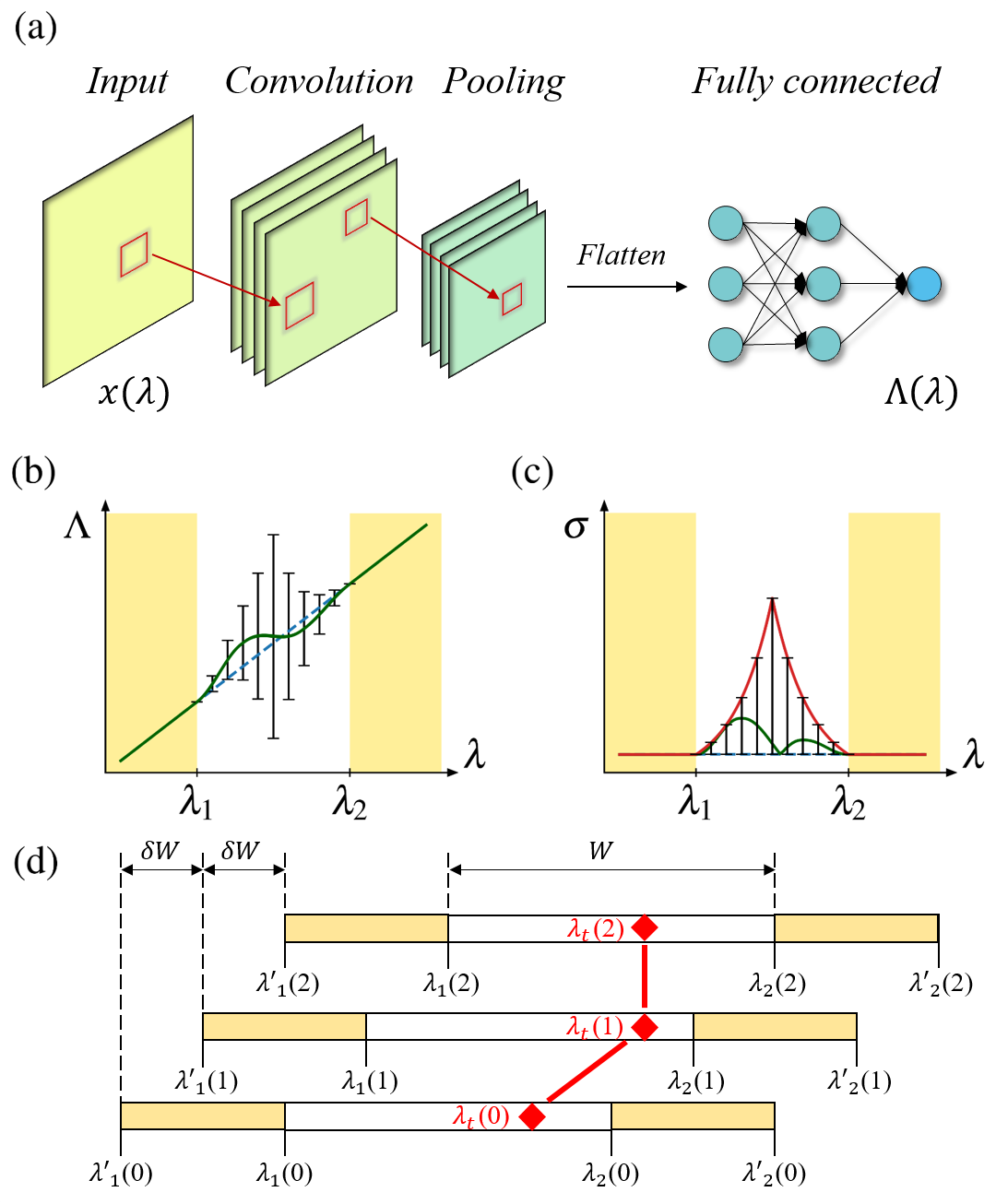}
    \caption{
    (a) The algorithm scheme for the SSL approach proposed in this paper for the identification of the PTP. The input features ($\{x(\lambda)\}$, say the ToF image) of a given system parameter ($\lambda$) will be transformed to the output, $\Lambda(\lambda)$, through a CNN. In (b), we show the training regions with yellow background and the test region with white background, where results obtained after repeating many trials with different initial random seeds. Green line indicates the the mean value, $\bar{\Lambda}(\lambda)$ and the error bar stands for the variance of the predicted results, $\sigma_{fluc}$. The diagonal straight line is a guidance for eye. In (c), we show the obtained $\sigma_{mean}$ and $\sigma_{fluc}$ in the test region (see Eq. (\ref{eq:sigma2})). The position of the maximum $\sigma_{fluc}$ is a candidate of the PTP.
    (d) shows a scheme diagram for the PTP candidates ($\lambda_t(n)$), red diamond) for different training/test regions. $W$ and $\delta W$ are the window size and the shift of test regions. See the text for more details
    }
    \label{fig:scheme_SSL.png}
\end{figure}

When a SSL model is trained, we apply it to predict an output, $\Lambda(\lambda)$, for a given system parameter, $\lambda$, in the test region. If the model is trained with 100\% accuracy, one should expect $\Lambda(\lambda)=\lambda$ in the test region (see the diagonal straight line in Fig. \ref{fig:scheme_SSL.png}(b)). However, since such a machine learning result is just a simulation, not the exact functional relationship, the statistical deviation between the known/controlled system parameter ($\lambda$) and its predicted value ($\Lambda$) should become the signature of the PTP, where the two phases are competing with each other and hence generate the largest fluctuations for the simulations. 

To implement the above physical concept in a realistic calculation, we could repeat the training process for $N_{tr}$ times with different random initial configurations of the neural networks, so that we could separate the deviation of the predicted mean value from the actual result and its statistical variation:
\begin{eqnarray}
\sigma^2_{tot}(\lambda)&=&\frac{1}{N_{tr}}\sum_{i=1}^{N_{tr}}|\Lambda_{i}(\lambda)-\lambda|^2
\nonumber\\
&=&\left|\bar{\Lambda}(\lambda)-\lambda\right|^2+\frac{1}{N_{tr}}\sum_{i=1}^{N_{tr}}|\Lambda_{i}(\lambda)-\bar{\Lambda}(\lambda)|^2
\nonumber\\
&\equiv &\sigma^2_{mean}+\sigma_{fluc}^2
\label{eq:sigma2}
\end{eqnarray}
where $\sigma_{mean}^2$ and $\sigma_{fluc}^2$ are two types of deviations, contributed independently from the mean value of the prediction ($\bar{\Lambda}(\lambda)$) and from the statistical fluctuations respectively. The former is a signature of the model performance and, in principle, could be reduced further if the model is better trained. On the other hand, the latter originates from the competition between the two phases: it is almost zero when deep in the parameter region of Phase $A$ or $B$, but become finite near the PTP (if only it exists in the test region). See Fig. \ref{fig:scheme_SSL.png}(b) and (c) for schematic description of the possible results. 

Different from the conventional SL approach mentioned in Sec. \ref{sec:supervised approach}, the  statistical fluctuation, $\sigma_{fluc}$ reflects the physical properties at the PTP and hence is not sensitive to the changes of training region. We will show that this method could be also applied to detect the existence of a phase transition even without any theoretical guidance. In the rest of this paper, we will show the calculation results of several known topological models and then provide further discussion in Sec. \ref{sec:discussion}. 

\subsection{Signature of Phase Transitions (II): Independence of the Training Region}

Besides of the largest deviation of the predicted results from the known system parameters, in our SSL approach, we also have to exam if the obtained PTP candidate (denoted by $\lambda_t$) is independent of the choice of the training/test region. Below we propose a systematic five-step process for this methodology.

Step I: We first choose an arbitrary training region in the system parameter space, $[\lambda_1'(0),\lambda_1(0)]$ and $[\lambda_2(0),\lambda_2'(0)]$ ($\lambda_2(0)>\lambda_1(0)$, see Fig. \ref{fig:scheme_SSL.png}(b) and (c)), where the training data are assumed randomly and equally distributed. The test region is then defined to be $(\lambda_1(0),\lambda_2(0))$, where we could apply the SSL model to get the PTP candidates, $\Lambda_t(0)$. 

Step II: Similarly, we could then define a series of the training regions by shifting all the region one step ($\delta W$) a time, i.e. $[\lambda_1'(n),\lambda_1(n)]$ and $[\lambda_2(n),\lambda_2'(n)]$, where $\lambda_{1/2}(n)\equiv \lambda_{1/2}+n\delta W$ and $\lambda_{1/2}'(n)\equiv \lambda_{1/2}'+n\delta W$, see Fig. \ref{fig:scheme_SSL.png}(d). As a result, the test region will be also shifted to accordingly to $(\lambda_1(n),\lambda_2(n))$ by keeping the test window size, $W\equiv \lambda_2(n)-\lambda_1(n)$ the same. We also require the density of the random chosen training/test data in these new training/test regions are also kept the same for consistence.  

Step III: We could also get the PTP candidates, $\lambda_t(n)$, for each region, see Fig. \ref{fig:scheme_SSL.png}(d). Here we emphasize that such candidates could be obtained either by equating the probability to find these two phases in the conventional SL approach, or by calculating the largest deviation of the predicted system parameter from its actual value, $\sigma_{fluc}$, as stated in Sec. \ref{sec:self-supervised learning} for the SLL approach.

Step IV: Since the PTP candidates, $\lambda_t(n)$, should be pinned at $\lambda_c$, if only the latter is covered by the test regions in above series of measurement. Therefore we can calculate the standard deviation, $\sigma_p^{(j)}$, to investigate how a series of these candidates, $\lambda_t(j)\cdots\lambda_t(j+N_p-1)$, are gathered around themselves around their average value, $\bar{\lambda}_p(j)$. More precisely, we can define the following three quantities
\begin{eqnarray}
\bar{\lambda}_p^{(j)}&\equiv & \frac{1}{N_p}\sum_{n=0}^{N_p-1}\lambda_t(j+n) 
\label{eq:lambda_p}
\\
\left[\sigma_p^{(j)}\right]^2 &\equiv & \frac{1}{N_p}\sum_{n=0}^{N_p-1}|\lambda_t(j+n)-\bar{\lambda}_t(j)|^2
\label{eq:sigma_p}
\\
C_f^{(j)} &\equiv & 1-\frac{\sigma_p^{(j)}}{\sqrt{\sigma_p^{(j)2}+ \delta W^2}}.
\label{eq:C_f}
\end{eqnarray}
Here $N_p\equiv\textrm{Int}[W/2\delta W]$ is the number of successive PTP candidates to be considered in order to make sure that most representative results within the test window are included. In Eq. (\ref{eq:C_f}), we define the "confidence" ($C_f^{(j)}$) as a function of the standard deviation so that the confidence is closer to one if these subsequent PTP candidates are congregated to each other (i.e. $\sigma_p^{(j)}\ll W$), and becomes much smaller than one if they are diversified (i.e. $\sigma_p^{(j)}\sim \delta W$).

Step V: Finally, after calculating the confidence for different values of $j$, we define the final confidence to be their maximum value, $C_f\equiv\text{Max}(C_f^{(j)})$, where the corresponding value of candidates average, $\lambda_p$, could be also obtained. We could then conclude that the value of $\lambda_p$ should be the most possible position of the PTP with a confidence $C_f$. This result is more/less reliable if the corresponding $C_f$ is larger/smaller.

Note that, using this approach, we do not have to assume the existence of a PTP inside the test region, and could apply this approach to find the possible new PTP from the experimental data directly even without experimental guidance. At the same time, we could also utilize this definition to investigate some exactly solvable models and to compare results obtained by conventional SL approach and our SSL approach, where the way to get the PTP candidates are different (the former is by equating the predicted probabilities of the two phases, while the latter if by calculating the maximum value of statistical fluctuations). Below we will show how this new methodology works for some exactly solvable models in details.

\section{Application to Intrinsic Topological Systems} 
\label{sec:Intrisic Topological}

Here we first apply the SSL approach to investigate the topological phase transition of intrinsic topological systems, including 1D Su-Schrieffer–Heeger model and 2D Haldane model. We will see that although the topological properties, dimensions, and related mechanism of these two models are very different, the associated topological phase transition could be determined equally well in our SSL approach. We will also show why the conventional SL approach could not provide satisfactory results. 

\subsection{1D SSH Model}
\label{sec:SSH}

1D Su-Schrieffer–Heeger (SSH) model \cite{Intro_SSH} is known to be the simplest model with a topological order. It describes a long chain of atoms with alternating two different bonds, and hence has the following single particle Hamiltonian:
\begin{equation}
\hat{H}_{SSH} =\sum_{i=1}^{L_x/2}(v\hat{c}^\dagger_{2i-1}\hat{c}_{2i}+w\hat{c}^\dagger_{2i}\hat{c}_{2i+1})+h.c.,
\end{equation}
where $v$ and $w$ are the tunnelling amplitudes between alternating bonds. $\hat{c}_{2i}$ and $\hat{c}_{2i-1}$ are the electron field operators at even and odd sites respectively. Here we just consider the case with even number of total lattice site, $L_x$ for simplicity. Due to the competition of the two bonds, the system has two kinds of dimerization mode, corresponding to a topological phase as $v<w$ and a trivial phase as $v>w$.

\begin{figure}[b]
    \centering
    \includegraphics[width=0.48\textwidth]{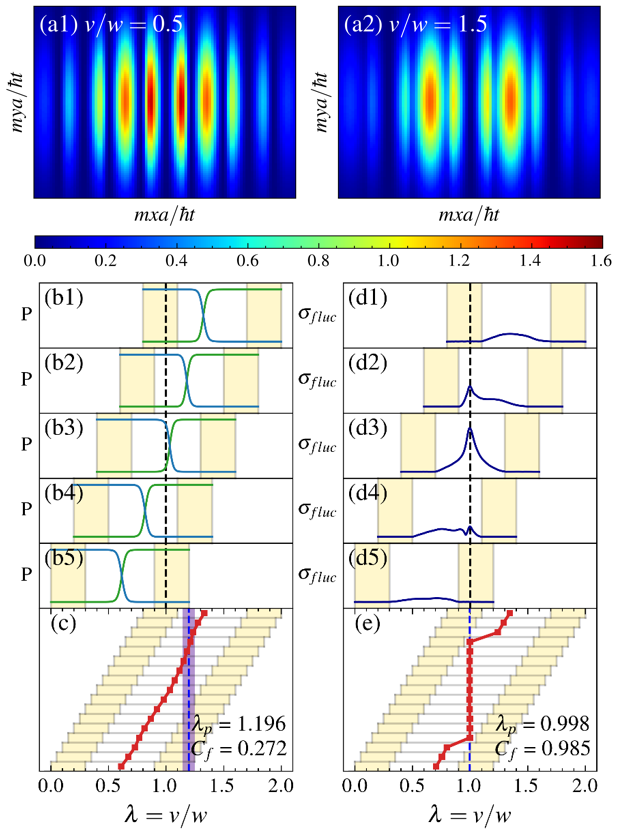}
    \caption{
    (a1) and (a2) are typical ToF images obtained from the SSH model with $L_x=200$ for $v/w=0.5$ (topological) and $v/w=1.5$ (trivial) phases respectively.
    (b1)-(b5) show the calculated probabilities for the test data to be in Phase I (topological) or Phase II (trivial) by the conventional SL approach. Results of different training/test regions (yellow/white background) are shown together for comparison (see Sec. \ref{sec:self-supervised learning}), with the vertical dashed line being the position of the true phase transition point. Here we choose the test data window $W=0.6$ and $\delta W=0.05$ (see Fig. \ref{fig:scheme_SSL.png}(d)). 
    (c) Red lines represent the obtained PTP candidates, $\lambda_t(n)$, as a function of different training/test regions for the conventional SL approach. The largest confidence ($C_f=0.272$) is denoted by the vertical dashed line at $\lambda_p=1.196$. Purple region indicate the related standard deviation ($\sigma_p$). (d1)-(d5) show the calculated fluctuations ($\sigma_{fluc}^{(j)}$, blue lines) of the predicted system parameters from the known results in the SSL approach. All other parameters are the same as in the corresponding figures in (b1)-(b5). (e) shows the obtained PTP candidates (red lines) in the same form as (b). The position of the largest confidence ($C_f=0.985$) is at $\lambda_p=0.998$.
    }
    \label{fig:Prediction_SSH.png}
 \end{figure}

In order to compare the results of conventional SL approach and the SSL approach proposed here, we use the ToF images (see Appendix \ref{sec:input features} for details) as the input features (see Figs. \ref{fig:Prediction_SSH.png}(a1) and (a2)) and systematically train these models in different training and test regions, as shown in Figs. \ref{fig:Prediction_SSH.png}.
One could see that for the former case ((b1)-(b5)), the predicted probabilities to be either phase do strongly depend on the the training region as mentioned in Sec. \ref{sec:supervised approach}. As a result, the obtained $\lambda_t(n)$ also changes as the training/test region shift as shown in (c). The PTP with the one of the largest confidence is at $\lambda_p=1.196$, which is 20\% off the known result, and the associated confidence is also pretty low ($C_f=0.272)$. As a result, we find that the conventional SL approach could not provide a reliable prediction of the PTP, unless a specific training region is chosen according to the pre-known position of the PTP.

On the other hand, from the results obtained by the SSL approach in (d1)-(d5), the largest statistical deviation ($\sigma_{fluc}$) of the predicted system parameter ($\Lambda$) from the known system parameter ($\lambda$) could make the candidates of PTPs ($\lambda_t$) almost pinned exactly at the true phase transition point, $\lambda_p=0.998\sim\lambda_c=1$ with a very high confidence ($C_f=0.985$), if only the latter is within the test region.

{
On the other hand, such a $\lambda_p$ changes accordingly to the changes of training region if the true phase transition point, $\lambda_c$, is not included inside the test region. As a result, as summarized in (d), the position of the predicted transition point, $\lambda_p$ is fixed almost like a constant if it is covered by the test region. Therefore, one could see the results predicted by self-supervised learning approach are much more robust than the conventional supervised approach even without knowing the existence of the phase transition or not.
}

\subsection{2D Haldane Model}
\label{sec:Haldane}

\begin{figure}[h]
    \centering
    \includegraphics[width=0.48\textwidth]{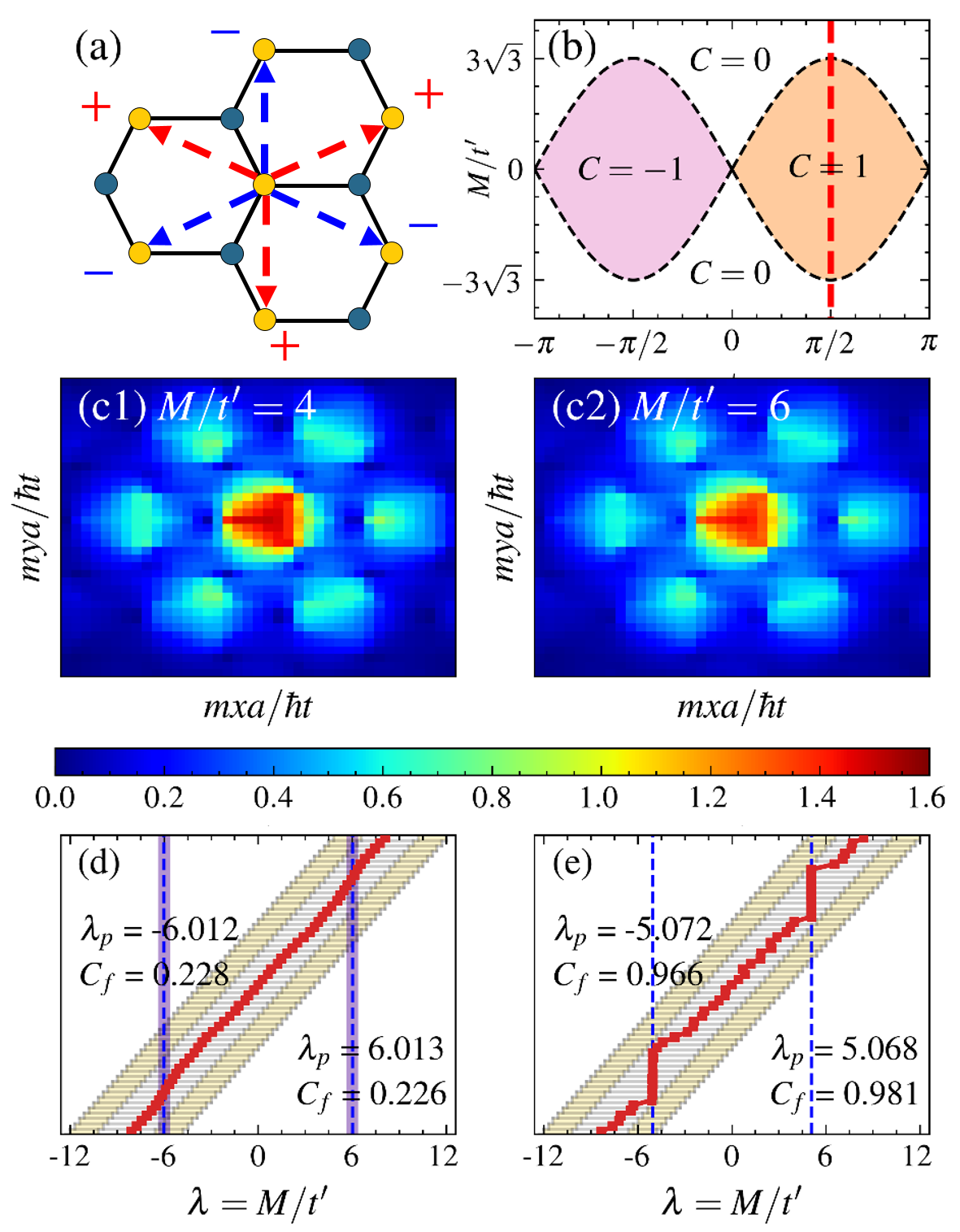}
    \caption{
    (a) The schematic lattice structure and model representation of the 2D Haldane model \cite{Intro_Haldane}. Different sublattice have different on-site energy as indicated by the blue and yellow circles. The red/blue dashed arrows indicate the the next nearest neighboring site tunneling ($t'$) with positive/negative phase, $\phi$. 
    (b) The topological phase diagram in terms of $phi$ and the sublattice on-site energy difference, $M/t'$. Regions of different Chern numbers ($C$) are also shown together. 
    (c1) and (c2) are typical ToF images obtained from the 2D Haldane model for $M/t'=4$ (topological) and $M/t'=6$ (trivial) phases respectively. Here we set $t'/t=1/3$.
    (d) and (e) are the obtained phase transition point candidates, $\lambda_t(n)$, for conventional SL approach and our SSL approach respectively. All other notations are the same as the corresponding panels in Figs. \ref{fig:Prediction_SSH.png}(c) and (e).
    }
    \label{fig:Prediction_Haldane.png}
 \end{figure}

Now we want to investigate if the SL approach could also be applied to another important, 2D Haldane model \cite{Intro_Haldane}, which has been experimentally realized and measured in the systems of ultracold atoms \cite{Exp_Haldane}. The 2D Haldane model is known to be a propo-type of the anomalous quantum Hall effect with non-zero Chern numbers. As shown in Fig. \ref{fig:Prediction_Haldane.png}(a), its Hamiltonian is defined on a honeycomb lattice:
\begin{eqnarray}
    \hat{H}_{Haldane} &=&\sum_{j}(-1)^jM\hat{c}^\dagger_j\hat{c}_j^{}-\sum_{\left<j,j'\right>}t\hat{c}^\dagger_j\hat{c}_{j'}^{}
\nonumber\\    
    &&-\sum_{\left<\left< j,j'\right>\right>}t'e^{i\phi_{jj'}}\hat{c}^\dagger_j\hat{c}^{}_{j'},
\end{eqnarray}
where $M$ represents the on-site energy difference, $t$ and $t'$ represent the nearest neighboring hopping and the next nearest neighboring hopping respectively. $\phi_{jj'}$ are of constant value but with opposite signs for different direction, arising from the staggered magnetic field and the Aharonov–Bohm effect. In Fig. \ref{fig:Prediction_Haldane.png}(b), we show its topological phase diagram with $60\times3$ unit cells and $t=1$, where the topological orders exist for $|M/t'|<3\sqrt{3}sin\phi$. Taking $t'/t=1/3$ and $\phi=\pi/2$ (the red vertical dashed line), we show the typical ToF images for a topological/trivial phase in Figs \ref{fig:Prediction_Haldane.png}(c1)/(c2). As one can see, their differences seem too small to be distinguishable by human eyes. It is therefore instructive to see if any machine learning approach could work much better.

In Figs. \ref{fig:Prediction_Haldane.png}(d) and (e), we show the PTP candidates calculated by the conventional SL approach and by the SSL approach. All the figure notations stand for the same meaning as those for the 1D SSH model in Figs. \ref{fig:Prediction_SSH.png}(c) and (e), unless there should be two PTPs in the parameter range we consider ($-12<\lambda=M/t'<12$).
As we could see from Fig. \ref{fig:Prediction_Haldane.png}(d), both PTPs predicted by the conventional SL approach have pretty low confidence ($C_f<0.23$) and the obtained values are about 15.7\% deviated from the exact results ($|\lambda=M/t'|=3\sqrt{3}=5.196$.
On the other hand, for the same training/test regions, the results provided by our SL approach works much better: the predicted position of the PTP is only 2.4\% away from the exact result (note that a finite deviation must exist due to the finite size effect), and the obtained confidences are also very high ($C_f>0.96$) for both transitions. This example also demonstrate the possibility to apply our SSL approach for the study of two phase transition points, even without setting any specific training/test regions beforehand.

\section{Application to Symmetry Protected Topological Systems}
\label{sec:Symmetry Topological}
\subsection{1D Kitaev Chain}
\label{sec:Kitaev model}

\begin{figure}[h]
    \centering
    \includegraphics[width=0.48\textwidth]{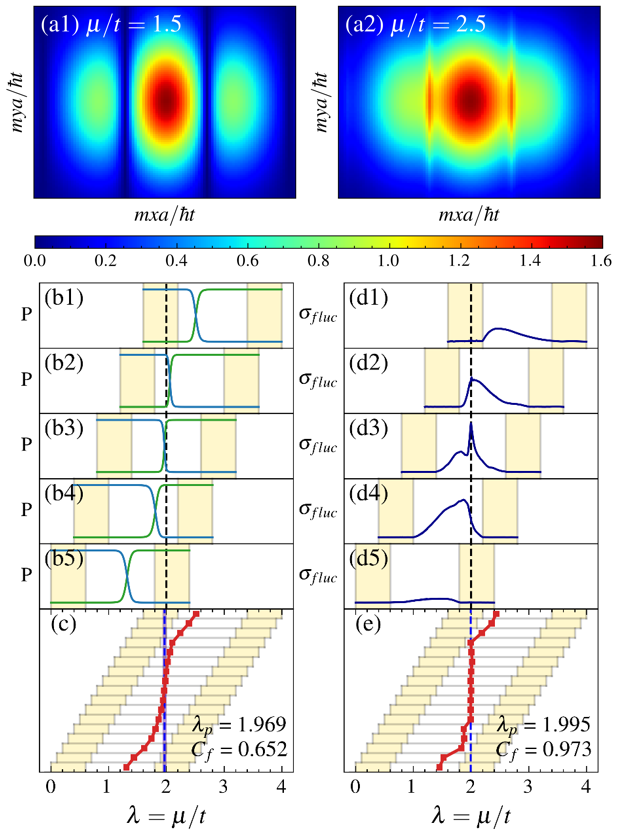}
    \caption{
    (a1) and (a2) show the typical ToF images obtained from the Kitaev chain for $\mu/t=1.5$ (topological) and $\mu/t=2.5$ (trivial) phases respectively. (b1)-(b5) show the calculated probabilities for the test data to be in Phase I (topological) or Phase II (trivial) by the conventional SL approach. Results of different training/test regions (yellow/white background) are shown together for comparison with the vertical dashed line being the position of the true PTP. Here we choose the test data window $W=1.2$ and $\delta W=0.1$ (see Fig. \ref{fig:scheme_SSL.png}(d)). (c) shows the position of these PTP candidates for various training/test regions. (d1)-(d5) and (e) are calculated fluctuations ($\sigma_p^{(j)}$) and the obtained PTP candidates by the SSL approach. The predicted PTP position, $\lambda_p$, and the corresponding confidence, $C_f$, are also shown in (c) and (e) respectively. All other notations are the same as the corresponding panels in Fig. \ref{fig:Prediction_SSH.png}.
    }
    \label{fig:Prediction_Kitaev.png}
 \end{figure}

In addition to the intrinsic topological systems, it is also important to investigate how the conventional SL approach and the SSL approach work for the symmetry protected topological systems. After all, the later depends on the presence of many-body order parameters and may not be easily observed unless in a very low temperature regime. Here we first consider the simplest topological superconductor, 1D Kitaev Chain \cite{Intro_Kitaev}, which has Majorana zero modes in the topological regime and has the following Hamiltonian:
\begin{equation}
\hat{H}_{Kitaev}=\sum_{n}(-\mu \hat{c}^\dagger_{n}c_{n}
-t\hat{c}^\dagger_{n+1}c_{n}
+\Delta \hat{c}^\dagger_{n+1}\hat{c}^\dagger_{n})+h.c.
\end{equation}
where $\mu$ is the chemical potential, $t$ is the tunnelling amplitude and $\Delta$ is the real-space superconducting pairing, which can be implemented by proximity effect \cite{Nanowire_Experiment}. The competition between the former two quantities (in the presence of a finite $\Delta$) leads to two different edge modes, unpaired Majorana mode (topological phase) as $|\mu|<2t$ and paired Majorana mode (trivial phase) as $|\mu|>2t$. The corresponding ToF images (input features) are shown in Figs \ref{fig:Prediction_Kitaev.png}(a1) and (a2) respectively with the system size $L_x=200$ and $\Delta/t=1$. 

In Figs. \ref{fig:Prediction_Kitaev.png}(b1)-(b5), 
we show the predicted probabilities to be either a topological phase or a trivial phase for the conventional SL approach. The yellow/white backgrounds indicate the training/test regions as shown in Fig. \ref{fig:Prediction_SSH.png}(b1)-(b5). We could see the obtained $\lambda_t(n)$ (by equating the two probabilities) also changes as the training/test region shift, summarized in (c). The most possible phase transition point, $\lambda_p=\mu/t$, is 1.969, which is very close to the theoretical value, 2, while the confidence ($C_f$) is upto 0.652. The accuracy of the PTP position and the confidence is significantly higher than those obtained by 1D SSH model or 2D Haldane Model. This may imply that the conventional SL approach may work better for symmetry protected topological phases.

However, when we apply the SSL approach to calculate the PTP candidates and the associated confidence, as shown in (d1)-(d5) and summarized in (e), our SSL approach can still work much better with the more precise prediction ($\lambda_p=1.995$) and higher confidence ($C_f=0.973$). This result therefore still suggest that the SSL approach is a better method for the identification of topological phase transitions.

\subsection{2D P-Wave Superfluid}
\label{sec:p-wave model}

\begin{figure}[h]
    \centering
    \includegraphics[width=0.48\textwidth]{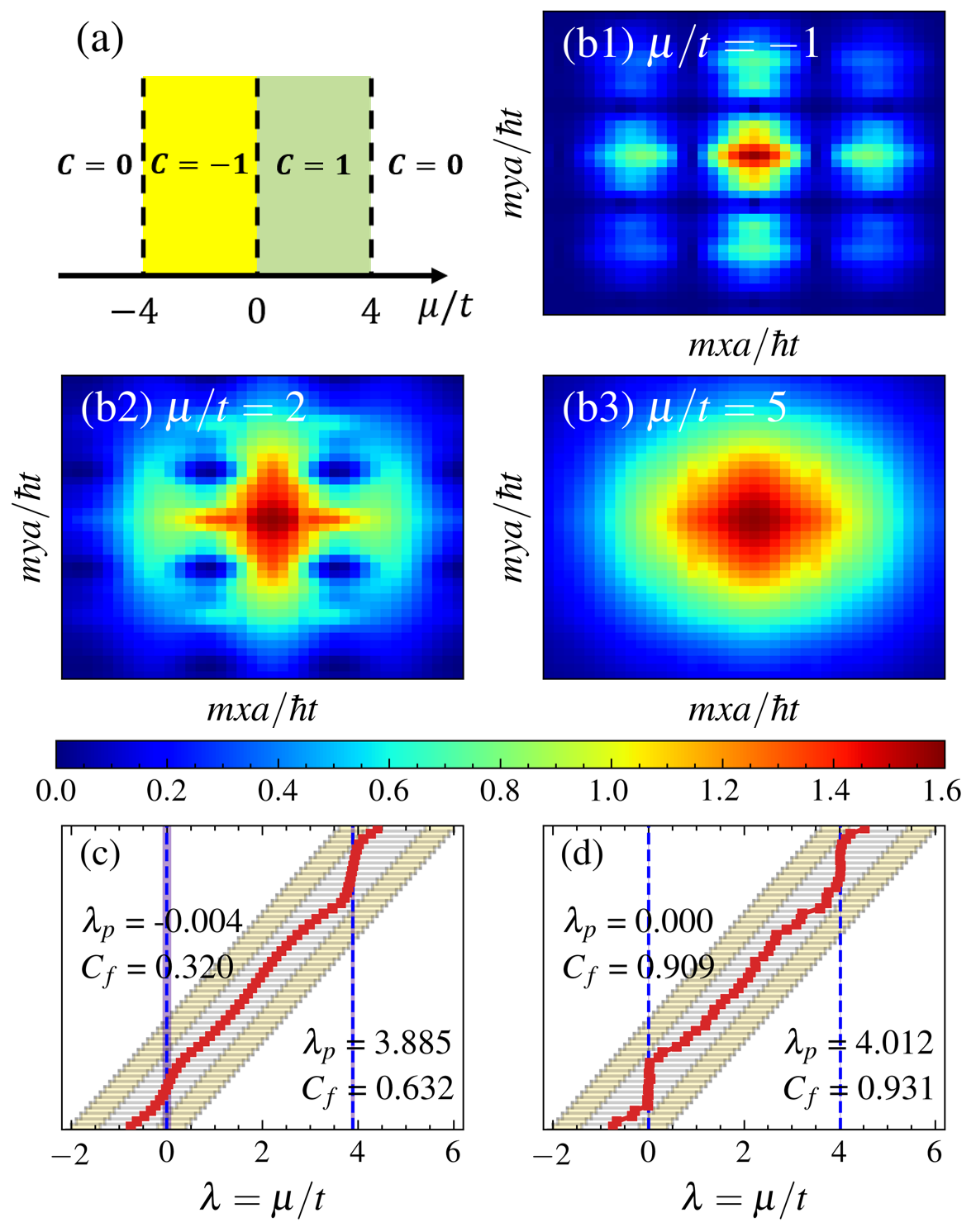}
    \caption{
    (a) The topological phase diagram in terms of $\mu/t$ for a 2D $p$-wave superfluid. Regions of different Chern numbers ($C$) are shown together. 
    (b1)-(b3) are the typical ToF images obtained from  for $\mu/t=-1$ (C=-1), $\mu/t=2$ (C=1) and $\mu/t=5$ (C=0) respectively. Here we set $\Delta/t=0.5$. This result is obtained by a finite size system with $L_x=60$ and $L_y=4$.
    (c) and (d) show the obtained PTP candidates, $\lambda_t(n)$, for conventional SL approach and our SSL approach respectively. All other notations are the same as the corresponding panels in Figs. \ref{fig:Prediction_SSH.png}(c) and (e).}
    \label{fig:Prediction_PWSC.png}
\end{figure}

Apart from 1D Kitaev chain, here we also investigate the possibility to identify another symmetry-protected topological phase, 2D $p$-wave superfluid \cite{Intro_pwave_1,Intro_pwave_2}, which could have a Majorana zero mode in vortices. The system Hamiltonian in a 2D square lattice could be given as following:
\begin{equation}
\begin{aligned}
\hat{H}_{p-wave}=\sum_{n,m}&(-\mu \hat{c}^\dagger_{n,m}\hat{c}_{n,m}
-t\hat{c}^\dagger_{n+1,m}\hat{c}_{n,m}-t\hat{c}^\dagger_{n,m+1}\hat{c}_{n,m}\\
&+\Delta \hat{c}^\dagger_{n+1,m}\hat{c}^\dagger_{n,m}+i\Delta \hat{c}^\dagger_{n,m+1}\hat{c}^\dagger_{n,m})+h.c.
\end{aligned}
\end{equation}
which has different gauge phase for pairing between two fermions in the $x$-axis and in the $y$-axis. It is known that such a $p$-wave superfluid breaks the time-reversal symmetry and hence has a chiral edge state at the boundary. In Fig. \ref{fig:Prediction_PWSC.png}(a), we show the topological phase diagram, where the topological states appear as $|\mu|<4t$ with opposite Chern number (or chiralities of the edge states) for different signs of $\mu$. In Figs. \ref{fig:Prediction_PWSC.png}(b1)-(b3), we show the typical ToF images for $\mu/t=-1$, 2, and 5, corresponding to topological phases of $C=-1$, 1, and 0 respectively. One could see that these features are quiet different from each other. As a result, one may expect that the phase transition boundary could be easily identified by conventional SL approach.

In Figs \ref{fig:Prediction_Kitaev.png}(c), we show the obtained results of the PTP candidates for the series of training/test regions. As on could see that the predicted position near $\mu=0$ is better predicted than the other one near $\mu/t=4$, while the confidence of the former is much less than the later, which is still below 0.65. For the results obtained by the SSL approach shown in Figs \ref{fig:Prediction_Kitaev.png}(d), on the other hand, we obtain much better results for both transition points (error is within 1\%) and very high confidence ($C_f>0.9$). Therefore, we still conclude that although the conventional SL approach may be applied to the identification of the phase transition boundary in the 2D $p$-wave case, it is still less effective compared to the SSL approach we proposed here.
\begin{figure}[h]
    \centering
    \includegraphics[width=0.48\textwidth]{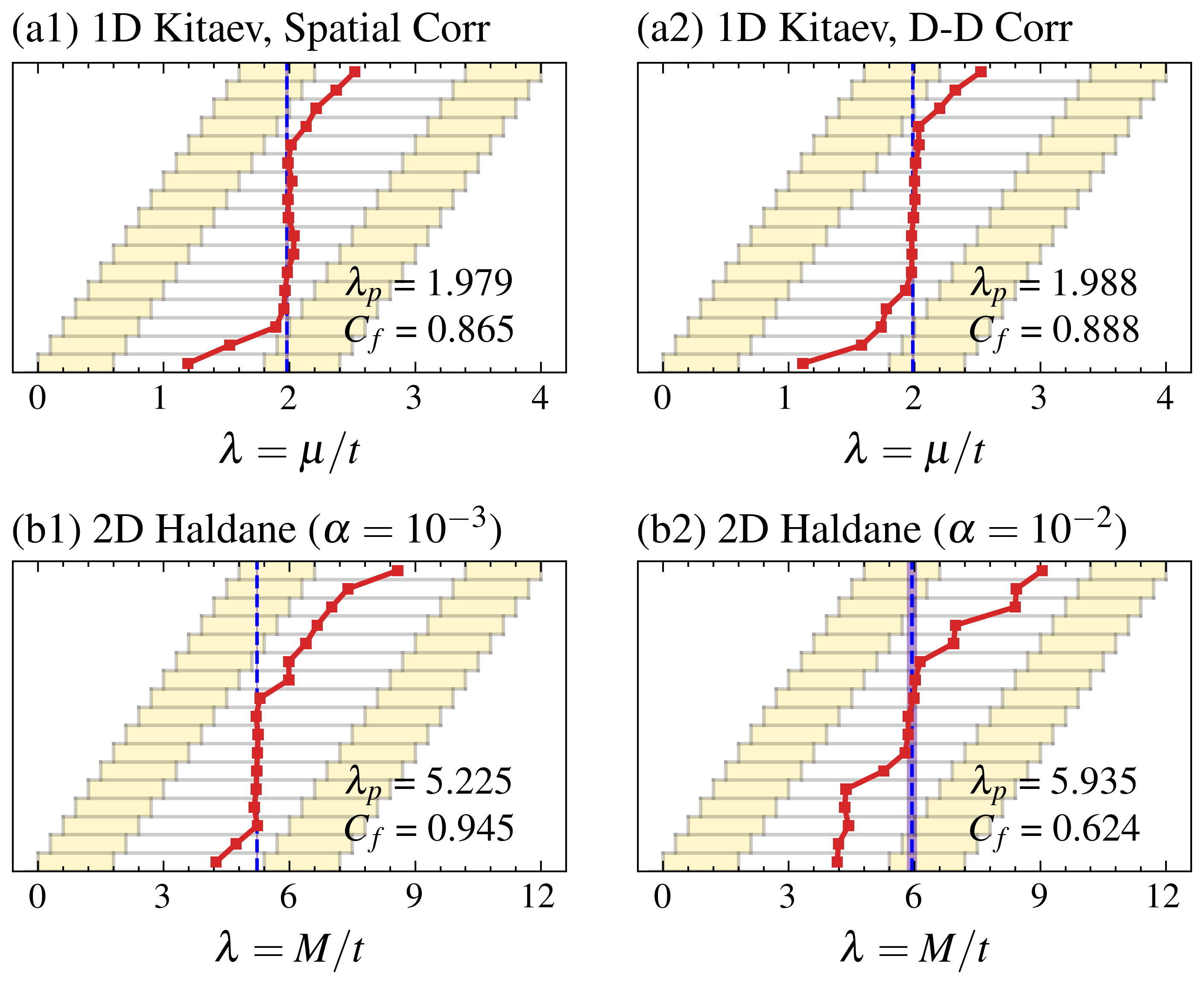}
    \caption{
    (a1) and (a2) are the PTP candidates, $\lambda_t(n)$, obtained by the SSL approach with input features as the spatial correlation function and the density-density correlation function respectively of a 1D Kitaev chain Model. The correct PTP is at $\lambda_p=2.0$. All other notations and parameters are the same as Fig. \ref{fig:Prediction_Kitaev.png}(e). (b1) and (b2) are the PTP candidates of the 2D Haldane model with two different finite harmonic potential for $\alpha=mw^2a^2/t=10^{-3}$ and $10^{-2}$ respectively (see the text). The correct PTP is at $\lambda_p=5.196$. All other notations and parameters are the same as Fig. \ref{fig:Prediction_Haldane.png}(e).
    }
    \label{fig:Prediction_Other.png}
 \end{figure}
\section{Discussion}
\label{sec:discussion}

\subsection{Spatial Correlation Function and Density-Density Correlation Function as Input Features}

Throughout this paper, we have used the time-of-flight (ToF) images as the major input features for our machine learning model. Although ToF images can be easily obtained in the systems of ultracold atoms, these data are hardly accessible for other condensed matter systems. Therefore, it will be more instructive to investigate if our conclusions could be retained when using other experimentally measurable features, say spatial correlation function or spatial density-density correlation function. In other words, it is important to check if our SSL approach could be generalized to other types of experiments for the precise and robust identification of topological phase transitions.   

In Figs. \ref{fig:Prediction_Other.png}(a1) and (a2), we show the calculated PTP candidates obtained respectively by using equal time spatial correlation function, $\langle \hat{c}^\dagger_i \hat{c}^{}_j\rangle/\sqrt{\langle \hat{n}_i\rangle \langle \hat{n}_j\rangle}$, and spatial density-density correlation function, $\langle \hat{n}_i \hat{n}_j\rangle/\langle \hat{n}_i\rangle \langle \hat{n}_j\rangle$, as the input features for the 1D Kitaev model as an example. Here we have normalize them by the mean value of the onsite density, $\hat{n}_i=\hat{c}^\dagger_i \hat{c}^{}_i$. We could see that the predicted phase transition points are still very accurate and with a reasonably high confidence. We also have applied conventional SL approach for the same calculation but did not get satisfactory results (not shown here), similar to the case of ToF images shown in Fig. \ref{fig:Prediction_SSH.png}(c). Therefore, it is reasonable to conclude that the SSL approach to identify the possible phase transition points can be directly applied to other types of experimental measurements.

\subsection{Effects of the Parabolic Trapping Potential}

In the theoretical models discussed above, we just consider uniform finite size systems with open boundary conditions. However, in the systems of ultracold atoms, there is always a harmonic trapping potential, $V_{trap}=\frac{1}{2}m\omega^2\vec{r}^2\equiv\frac{1}{2}\alpha t i^2$, generated by a magneto-optical trap to confine atoms in the vacuum. Here $m$ and $\omega$ are the mass of atoms and the effective confinement frequency. $\alpha\equiv m\omega^2a^2/t$ is the dimensionless energy scale with $a$ being the lattice constant and $t$ being the tunneling energy. It is therefore interesting and necessary to investigate how such an inhomogeneous potential may influence the identification of topological phase transition by the SSL approach. In Fig. \ref{fig:Prediction_Other.png}(b1) we show the obtained PTP candidates for the 2D Haldane model in the presence of a finite trapping potential. One could see that the predicted phase transition points are almost unchanged (within 3\% deviation from the results in Fig. \ref{fig:Prediction_Haldane.png}(e)) for a trapping strength, $\alpha=10^{-3}$, 
which is the same order as the experimental value of $^{40}K$ in Ref. \cite{Exp_Haldane,Exp_trap} with $\omega\sim 2\pi\times 22.5$ Hz and $a=532$ nm. However, when using a much larger trapping strength, say $\alpha=10^{-2}$, we could see that the predicted PTP will shift to a higher value with a much lower confidence ($C_f=0.624$). From the calculation above, we can see that the confidence level defined in our SSL approach is another important indicator to check if the predicted PTP is reliable or not, even when its existence has not be theoretically known before. In other words, we could use our SSL approach to distibuisg possible new topological phase transitions even without a priori knowledge about the underlying theories. 

\section{Conclusion}
\label{sec:conclusion}

In this paper, we have systematically investigated the problems of conventional supervised learning approach in the application of phase boundary prediction, and then proposed a self-supervised learning approach to resolve these problems. Using ToF images as input features of machine learning, we apply a self-supervised learning approach to investigate the phase transitions of various topological models, including 1D SSH model, 2D Haldane model, 1D Kitaev model and 2D $p$-wave superfluid model. We show that our SSL approach could always predict the topological phase transition precisely with very high confidence ($C_f>0.9$), much better than the results obtained for the conventional SL approach. This result could be understood from the simulation of a microscopic state function, which contains enough entanglement information to identify topological phases. We further show that these results could be also obtained by using spatial correlation function and density-density correlation function as input features, showing the generality and robustness of this approach even for other types of experimental measurements. Therefore, our SSL approach could be extended to other condensed matter systems for identifying possible new phase transitions solely based on the experimental data even without any a prior knowledge of the underlying theories.

\begin{acknowledgments}
We thank Chung-Yu Mou, Ming-Chiang Chung, Po-Chung Chen, and Po-Yao Chang for fruitful and valuable discussion. This work is supported by the Ministry of Science and Technology grant (MOST 107-2112-M-007-019-MY3) and by the Higher Education Sprout Project funded by the Ministry of Science and Technology and the Ministry of Education in Taiwan. We thank the National Center for Theoretical Sciences (Hsinchu, Taiwan) for providing full support. 
\end{acknowledgments}


\providecommand{\noopsort}[1]{}\providecommand{\singleletter}[1]{#1}%
%


\appendix
\section{Time-of-Flight Images as an input feature}
\label{sec:input features}

The ToF image is a special experimentally measurable quantities for the systems of ultracold atoms, obtained by measuring the column density distribution of a long time free expansion after released from all trapping potential. It is easy to see that, if the interaction between particles can be negligible when the density drops down during the expansion, and if the final atom cloud is much larger than the original size (long-time flight), the ToF image is equivalent to the project of momentum distribution integrated along the $z$ axis, and therefore can be calculated as following \cite{ToF_formula}
\begin{equation} 
    n_{ToF}(x,y) \propto \rho\left(\frac{m}{\hbar t}x,\frac{m}{\hbar t}y\right)\times
    \int_{-\infty}^{\infty}dz\left|\tilde{w}_0(m\mathbf{r}/\hbar t)\right|^2 
    \label{eq:1D_tof}
\end{equation}
where $\tilde{w}_{0}(\mathbf{k})$ is the Fourier transform of the lowest band Wannier function. $\rho(p_x,p_y)$ is the 2D momentum distribution in the $x-y$ plane {\it before} the trap is removed. Here we have used $p_x=\frac{m}{\hbar t}x$ to approximate the momentum in the $z$ direction by the final position based on the two assumptions mentioned above. $\rho(p_x,p_y)$ could be calculated from the real space correlation function as following:
\begin{eqnarray}
\rho\left(p_x,p_y\right)&=&\sum_{j_x,j_x'=1}^{L_x}\sum_{j_y,j'_y=1}^{L_y}
\langle  \hat{c}^\dagger_{j_x,j_y}\hat{c}^{}_{j'_x,j'_y}\rangle 
\nonumber\\
&&\times e^{ip_x(j_x-j_x')+ip_y(j_y-j_y')}
\label{eq:rho_p_2D}
\end{eqnarray}
where $\langle\cdots\rangle$ stands for the ground state expectation value. Note that in the momentum distribution shown above, we have just considered a 2D system, while this form could be easily applied to 1D system by taking $L_y=1$. Extension to 3D is also straightforward but will not be considered in the examples of this paper.

In Tab. \ref{tab:tofimage}, we list the resolution setting of ToF images simulated for each physical system of this paper. For simplicity, the lowest band Wannier function are approximated by Gaussian wavefunction with a width 0.141 times of lattice constant. 
\begin{table}[h] 
	\renewcommand{\arraystretch}{1.5}
	\tabcolsep=12pt   
	\centering
    \begin{tabular}{c c}
    \hline\hline
    System & Image Pixel\\
    \hline
	1D SSH & 100$\times$100\\ 
	2D Haldane & 40$\times$40\\
	1D Kitaev & 100$\times$100\\
	2D p-wave & 40$\times$40\\
	2D Haldane + Trap & 40$\times$40\\
    \hline\hline
    \end{tabular}
    \caption{
    The input number of pixels for the ToF images of each model discussed in the text. We use 500 images for training.
    }
    \label{tab:tofimage}
\end{table}
\section{Hyperparameters for Machine Learning Models}
\label{sec:ML parameters}

In Tab. \ref{tab:hyperparameters}, we list the hyperparameters of the machine learning algorithms mentioned in this paper. The construction of the supervised learning scheme is just a simple convolutional neural network (CNN), which includes one convolutional layer, one pooling layer and two fully-connected layers with scaled exponential linear units (SeLU) activation function. The outputs are normalized to the probabilities through softmax function and the loss function is chosen as the cross entropy (CE), where they are usually utilized in conventional classification. 
\begin{table*}[h] 
	\renewcommand{\arraystretch}{1.5}
	\tabcolsep=12pt   
	\centering
    \begin{tabular}{c c c c c}
    \hline\hline
    & \multicolumn{2}{c}{Supervised Learning} &\multicolumn{2}{c}{Self-supervised Learning}\\
    \hline
    Input &(2N, 2N, 1) &&(2N, 2N, 1)  \\
    Convolution &(2N, 2N, 10) &SeLU &(2N, 2N, 10) &SeLU\\
    Max-Pooling &(N, N, 10) &&(N, N, 10)\\
    Fully-Connected &(10N$^2$, 50) &SeLU &(10N$^2$, 50) &SeLU\\
    Output &(50, 2) &&(50, 1)\\
    \hline
	Optimizer&\multicolumn{2}{c}{Adam}&\multicolumn{2}{c}{Adam}\\
	Loss Function&\multicolumn{2}{c}{CE}&\multicolumn{2}{c}{MSE}\\
	Learning Rate&\multicolumn{2}{c}{$10^{-4}$}&\multicolumn{2}{c}{$10^{-4}$}\\
	Learning Step&\multicolumn{2}{c}{20000}&\multicolumn{2}{c}{20000}\\
    \hline\hline
    \end{tabular}
    \caption{The hyperparameters for supervised learning (SL) and self-supervised learning (SSL) we utilized in this paper. Here we take the input images with 2N$\times$2N pixels as an example.}
    \label{tab:hyperparameters}
\end{table*}

We note that, the implementation of our self-supervised learning scheme is almost exactly the same as the supervised learning one, and the only difference is the the number of neurons in the output and the loss function is chosen as mean square error (MSE). Besides, in order to let the learning algorithm more easily to be trained, we design a rescaling process for the self-supervised learning approach, which could be expressed in the following simple leaner form:
\begin{equation}
\Lambda(\lambda)=\frac{y(\lambda)-y(\lambda_i)}{y(\lambda_i)-y(\lambda_f)}\lambda_i+\frac{y(\lambda)-y(\lambda_i)}{y(\lambda_f)-y(\lambda_i)}\lambda_f
\end{equation}
where $y(\lambda)$ is the neural network's output, $\lambda_i$ and $\lambda_f$ are the two bounds of parameter set. This process could make sure that the predicted value are exactly equal to the groundtruth value at two bounds, i.e. $\Lambda(\lambda_i)=\lambda_i$ and $\Lambda(\lambda_f)=\lambda_f$.


\end{document}